\documentclass[aps,prd,reprint,showpacs,floatfix,footinbib,superscriptaddress,longbibliography]{revtex4-1}
\usepackage{orcidlink}
\usepackage{bbding}
\usepackage{amsmath}
\usepackage{amssymb}
\usepackage{latexsym}
\usepackage{graphics}
\usepackage{graphicx}
\usepackage{slashed}
\usepackage{color}
\usepackage{comment}
\usepackage{cancel}
\usepackage[normalem]{ulem}

\usepackage[disable,colorinlistoftodos,backgroundcolor=white,bordercolor=steelblue,linecolor=steelblue]{todonotes}

\usepackage{bm}

\definecolor{airforcelblue}{rgb}{0.36, 0.54, 0.66}
\definecolor{steelblue}{rgb}{0.27, 0.51, 0.71}
\definecolor{amber}{rgb}{1.0, 0.49, 0.0}

\makeatletter
\newsavebox\myboxA
\newsavebox\myboxB
\newlength\mylenA

\newcommand*\xoverline[2][0.75]{%
    \sbox{\myboxA}{$\m@th#2$}%
    \setbox\myboxB\null
    \ht\myboxB=\ht\myboxA%
    \dp\myboxB=\dp\myboxA%
    \wd\myboxB=#1\wd\myboxA
    \sbox\myboxB{$\m@th\overline{\copy\myboxB}$}
    \setlength\mylenA{\the\wd\myboxA}
    \addtolength\mylenA{-\the\wd\myboxB}%
    \ifdim\wd\myboxB<\wd\myboxA%
       \rlap{\hskip 0.5\mylenA\usebox\myboxB}{\usebox\myboxA}%
    \else
        \hskip -0.5\mylenA\rlap{\usebox\myboxA}{\hskip 0.5\mylenA\usebox\myboxB}%
    \fi}
\makeatother

\usepackage{booktabs,amsmath}
\usepackage{amssymb}
\usepackage{bm}
\usepackage{amsmath}
\usepackage{array}
\usepackage{mathtools}
\usepackage{amsthm}
\usepackage{newlfont}
\usepackage[utf8]{inputenc}
\usepackage{esdiff}
\usepackage{multirow}
\usepackage{graphics}
\usepackage{graphicx}
\usepackage{ae,aecompl,color}

\hypersetup{
    colorlinks,
    linkcolor={red},
    citecolor={cyan},
    urlcolor={blue!80!black}
}

\makeatletter
\DeclareRobustCommand*{\bfseries}{%
  \not@math@alphabet\bfseries\mathbf
  \fontseries\bfdefault\selectfont
  \boldmath
}
\makeatother

\begin{document}
\newcommand{\bd}{\begin{document}}
\newcommand{\ed}{\end{document}}
\newcommand{\bc}{\begin{center}}
\newcommand{\ec}{\end{center}}
\newcommand{\bfr}{\begin{flushright}}
\newcommand{\efr}{\end{flushright}}
\newcommand{\lt}{\left}
\newcommand{\rt}{\right}
\newcommand{\vs}{\vspace}
\newcommand{\hs}{\hspace}
\newcommand{\beq}{\begin{equation}}
\newcommand{\eeq}{\end{equation}}
\newcommand{\lb}{\linebreak}
\newcommand{\pb}{\pagebreak}
\newcommand{\mb}{\makebox}
\newcommand{\fb}{\framebox}
\newcommand{\mc}{\multicolumn}
\newcommand{\ben}{\begin{enumerate}}
\newcommand{\een}{\end{enumerate}}
\newcommand{\bit}{\begin{itemize}}
\newcommand{\eit}{\end{itemize}}
\newcommand{\un}{\underline}
\newcommand{\lefq}{\lefteqn}
\newcommand{\ba}{\begin{array}}
\newcommand{\ea}{\end{array}}
\newcommand{\beqa}{\begin{eqnarray}}
\newcommand{\eeqa}{\end{eqnarray}}
\newcommand{\beqas}{\begin{eqnarray*}}
\newcommand{\eeqas}{\end{eqnarray*}}
\newcommand{\bfg}{\begin{figure}}
\newcommand{\efg}{\end{figure}}
\newcommand{\bds}{\begin{displaymath}}
\newcommand{\eds}{\end{displaymath}}
\newcommand{\btb}{\begin{tabbing}}
\newcommand{\etb}{\end{tabbing}}
\newcommand{\para}{\parallel}
\newcommand{\pad}{\partial}
\newcommand{\nn}{\nonumber}
\newcommand{\la}{\leftarrow}
\newcommand{\ra}{\rightarrow}
\newcommand{\lgla}{\longleftarrow}
\newcommand{\lgra}{\longrightarrow}
\newcommand{\La}{\Leftarrow}\newcommand{\Ra}{\Rightarrow}
\newcommand{\Lra}{\Leftrightarrow}
\newcommand{\Lgla}{\Longleftarrow}
\newcommand{\Lgra}{\Longrightarrow}
\newcommand{\lan}{\langle}
\newcommand{\ran}{\rangle}
\renewcommand{\a}{\alpha}
\renewcommand{\b}{\beta}
\newcommand{\g}{\gamma}
\newcommand{\G}{\Gamma}
\renewcommand{\d}{\delta}
\newcommand{\eps}{\epsilon}
\newcommand{\Th}{\Theta}
\newcommand{\s}{\sigma}
\newcommand{\lam}{\lambda}
\newcommand{\D}{\Delta}
\newcommand{\vare}{\varepsilon}
\newcommand{\pr}{\prime}
\newcommand{\ro}{\rho}
\newcommand{\nab}{\nabla}
\newcommand{\m}{\mu}
\newcommand{\n}{\nu}
\newcommand{\Sg}{\Sigma}
\newcommand{\p}{\pi}
\newcommand{\R}{I\!\!R}
\newcommand{\om}{\omega}
\newcommand{\Om}{\Omega}
\newcommand{\ze}{\zeta}
\newcommand{\vart}{\vartheta}
\newcommand{\tri}{\triangle}
\newcommand{\f}{\frac}
\newcommand{\iny}{\infty}
\newcommand{\pro}{\propto}
\renewcommand{\arraystretch}{1.25}
\title{Quantum backflow of a Dirac fermion on a ring } 
%
%
\author{\textsc{Valentin Daniel Paccoia}\orcidlink{0000-0001-8799-6127}}
\affiliation{Dipartimento di Fisica e Geologia, Università degli Studi di Perugia, Via A. Pascoli, 06123, Perugia, Italy}
\email[Email:]{valentindaniel.paccoia@studenti.unipg.it}
\author{\textsc{Orlando Panella}\orcidlink{0000-0003-4262-894X}}
\affiliation{Istituto Nazionale di Fisica Nucleare, Sezione di Perugia, Via A.~Pascoli, I-06123 Perugia, Italy}
\email[({\bf Corresponding Author}) Email: ]{orlando.panella@pg.infn.it }
\author{\textsc{Pinaki Roy}\orcidlink{0000-0001-9842-2268}}
\affiliation{Atomic Molecular and Optical Physics Research Group, Advanced Institute of Materials Science,
Ton Duc Thang University, Ho Chi Minh City, Vietnam}
\email[Email:]{pinaki.roy@tdtu.edu.vn}
\affiliation{Faculty of Applied Sciences, Ton Duc Thang University, Ho Chi Minh City, Vietnam}

\date{\today}

\begin{abstract}
We study the quantum backflow problem of a relativistic charged Dirac fermion constrained to move on a ring of radius $R$. Using the relativistic current operator we compute the probability flux through a generic time interval to show emergence of quantum backflow. We also discuss the limiting case when the particle moves along a line.
\end{abstract}


\maketitle

\section{Introduction}
Quantum backflow (QB) is a phenomenon in which the probability density of a quantum mechanical particle may flow in a direction opposite to the momentum \cite{allcock1,Bracken_1994}. One of the most interesting results in this context is that there is a limit on the maximum amount of backflow. It essentially means that although backflow may increase with time, the increase is bounded by a dimensionless number $c_{bm}\approx 0.04$ \cite{Bracken_1994}. Later, improved estimates of the number $c_{bm}$ were found \cite{eveson,penz2005}. QB has been studied in the context of various nonrelativistic systems e.g., optical systems \cite{berry10,Eliezer:20}, elementary quantum systems \cite{yearsley2012}, particle in a constant  magnetic field \cite{strange}, correlated quantum states \cite{PhysRevResearch.2.033206}, many-particle systems \cite{Barbier2020:aa}, systems with a scattering potential \cite{PhysRevA.96.012112,cadamuro2017quantum}, noncommutative plane \cite{PhysRevA.102.062218}, open quantum systems \cite{Mousavi2020}, dissipative systems \cite{mousavi:2020aa}. Interestingly, quantum backflow has also been found to be related to nonclassicality \cite{doi:10.1142/S0219749916500325}, superoscillations \cite{muga2000}, decay of quasi stable systems \cite{PhysRevA.100.052101} and classically forbidden probability flow \cite{PhysRevA.99.043626}, transmitting defects \cite{vasconcelos2020quantum}, negative flow of probability \cite{doi:10.1119/10.0000856} etc. It is thus natural to inquire whether or not the concept of QB can be extended to relativistic systems which are by nature more complicated than the nonrelativistic ones. The answer to this question is in the affirmative and it was shown that, despite the localization problem of Dirac particles, the QB effect does take place for a Dirac particle described by the one particle Dirac equation \cite{Melloy1998:aa,su2018quantum,Ashfaque:2019aa,das2021relativistic}. 
It has also been shown that, for a free Dirac particle, the probability that flows in a direction opposite to the momentum in a time interval of length $T$ depends on a dimensionless parameter $\eps$ which is a function of mass $m$, velocity of light $c$, Planck's constant and $T$ \cite{Melloy1998:aa}. Later, backflow eigenfunctions were determined numerically over a range of values of $\eps$ and an explicit expression for relativistic backflow was found \cite{Ashfaque:2019aa}. Also, backflow of a free Dirac particle in a superposition consisting of various energies and helicities \cite{Su:2018aa}, in electron wave packets \cite{das2021relativistic} and for spin-orbit coupled particles \cite{mardonov2014interference} have been studied. Interestingly QB has also been studied recently within the context of Dirac equation, Maxwell equations and linearized gravity \cite{10.1088/1751-8121/ac65c1}.

Although there have been formulations of quantum backflow more amenable to experimental verification \cite{palmero,Miller2021experimentfriendly,Barbier2021experimentfriendly,PhysRevA.107.032204}, this effect, whether in the nonrelativistic domain or in the relativistic domain, is an as yet unobserved phenomenon. The main reason for the non observation of QB is that the net backflow through a point over a finite time interval is a very small quantity~\cite{Melloy1998:aa}. However, it has been shown  that, when a nonrelativistic particle is constrained to move on a ring~\cite{Goussev2020:ab}, the chances of observing quantum backflow increase manyfold as compared to when the particle moves on a line. In the present paper our aim is to 
examine whether or not this strategy works
for a relativistic quantum mechanical system. More precisely, we shall consider a Dirac fermion on a ring of radius $R$ in the presence as well as in the absence of a constant  magnetic field $B$ and for this system we shall compute the quantum backflow as a function of the aforementioned parameter $\eps$ and $B$. Finally, we shall also compare our results in the $c\ra \infty$ limit with those of the nonrelativistic backflow problem. 

The rest of the paper is organised as follows: in Sec.~\ref{Sec:problemsetup} we set up  the problem of the relativistic fermion on a circular ring of radius $R$; 
 Sec.~\ref{Sec:QBF} discusses the quantum backflow of the relativistic fermion;
 Sec.~\ref{Sec:numerical} discusses the full numerical analysis of the corresponding homogeneous Fredholm integral equation to which the quantum backflow problem on the ring is reduced. Finally Sec.~\ref{Sec:conclusions} presents the discussion and conclusions.

\section{Problem setup}
\label{Sec:problemsetup}
The Hamiltonian of a charged Dirac fermion constrained on a plane (a two-dimensional system) reads:
\begin{equation}
\label{hamiltonian0}
H= c \,\bm{\sigma}\cdot{\bm{P}} + mc^2\sigma^3,   
\end{equation}
where $\bm{\sigma}= (\sigma^1,\sigma^2)$, and $\sigma^i$, ($i=1,2,3$) are the Pauli matrices and the coupling to the electromagnetic field is introduced via the minimal prescription:
\begin{equation}
    \bm{P} = \bm{p} -\frac{q}{c} \bm{A}\, ,
\end{equation}
where $\bm{p}=-i\hbar\,(\partial_x,\partial_y)$ and $\bm{A}$ is the vector potential. 

In this work we will consider the dynamics of a charged fermion in a constant magnetic field directed along the $z$-axis, $\bm{B}= B\bm{e}_z$, perpendicular to the $(x,y)$ plane, which we describe in the so-called symmetric gauge with the following potential:
\begin{equation}
    \bm{A} = \left(-\frac{By}{2}, \frac{Bx}{2}\right)\,.
\end{equation}
The Hamiltonian in Eq.~\eqref{hamiltonian0} becomes:
\begin{equation}
  H = \left( \begin{array}{cc}
  mc^2& cP_-\\
  cP_+& -mc^2
  \end{array}\right), 
\end{equation}
where we have defined:
\begin{equation}
 P_{\pm}=P_x \pm iP_y   \, .
\label{ladderP}
\end{equation}
The operators $P_\pm$ are conveniently written in polar coordinates:
\begin{equation}
    P_\pm = - i\, e^{\pm i \varphi}\, \left[\left( \hbar\, \partial_r \pm i\, \frac{\hbar}{r}\,\partial_\varphi\right) \pm \frac{qBr}{2 c}\right].
\end{equation}

In the following we will be interested specifically in the quantum backflow of the Dirac fermion on the plane when it is constrained  to move on a circle of radius $R$. The circle is assumed to be centered at the origin $O=(0,0)$ of the $(x,y)$ coordinate system. Therefore, imposing the constraint $r=R$, the operators $P_\pm$ in Eq.~\eqref{ladderP} are given by:
\begin{equation}
  c\,P_\pm = - i\, \frac{\hbar c}{R} e^{\pm i \varphi}\, \left(  \pm i\, \partial_\varphi \pm \frac{qB R^2}{2\hbar c}\right)  \, .
\end{equation}
We now introduce the dimensionless quantity 
\begin{equation}
\label{betafactor}
    \beta= \frac{qBR^2}{2 \hbar c}. 
\end{equation}
This gives the flux $\Phi$  of the magnetic field across the circle of radius $R$, $\Phi=\pi B R^2$,   in units of the flux-quantum $\Phi_0 =hc/e$  (assuming the charge $q$ to be measured in units of $e$). The $P_\pm$ are then written as:
\begin{equation}
  c\,P_\pm = \pm \, \frac{\hbar c}{R} e^{\pm i \varphi}\, \left(  \partial_\varphi -i \beta \right)  \, ,
\end{equation}
so that the Hamiltonian in Eq.~\eqref{hamiltonian0} becomes:
\begin{equation}
\label{hamiltonian1D}
  H = \left(\begin{array}{cc}
  mc^2& -\frac{\hbar c}{R}e^{-i\varphi}\left( \partial_\varphi -i\beta\right)\\
  \frac{\hbar c}{R}e^{+i\varphi} \left(\partial_\varphi -i\beta\right)& -mc^2
  \end{array}\right). 
\end{equation}
The time dependent  Dirac equation with the 2-dimensional Hamiltonian in Eq.~\eqref{hamiltonian0}: 
\begin{equation}
\label{time-dep-Dirac}
i\hbar\frac{\partial \psi}{\partial t} = H\psi,   
\end{equation}
admits a continuity equation which, 
for a generic two-component  spinor state $\psi=(\chi_1,\chi_2)^T$, can be written as:
\begin{equation}
\label{continuity}
    \frac{\partial \rho}{\partial t} + \bm{\nabla} \cdot \bm{J} 
    = 0,
\end{equation}
where  the probability density is given by:
\begin{equation}
\label{density}
    \rho = \psi^\dagger \, \psi = \chi_1^* \chi_1^{\phantom{*}} + \chi_2^* \chi_2^{\phantom{*}}, 
\end{equation}
and the current $\bm{J}$ is
\begin{equation}
    \label{2dimcurrent} 
    \bm{J} = c\, \psi^\dagger \bm{\sigma}\, \psi\,.
\end{equation}
The explicit expression of the current in Eq.~\eqref{2dimcurrent} is the usual $U(1)$ current of the Dirac equation minimally coupled to the electromagnetic field. The two-dimensional time dependent Dirac equation in Eq.~\eqref{time-dep-Dirac} is written in manifestly covariant form by introducing a set of two-dimensional gamma matrices $\gamma^\mu$ as $\gamma^0=\sigma^3$, $\gamma^i=\sigma^3 \sigma^i$, ($i=1,2$) which satisfy the Clifford algebra. Then, a straightforward application of Noether's theorem to the global  phase symmetry provides the well known $U(1)$ current $j^\mu=\bar{\psi} \gamma^\mu \psi$  with $\bar{\psi} = \psi^\dagger \gamma^0$ which is conserved $\partial_\mu j^\mu=0$. This last condition, when written out explicitly  
in terms of its time and space components, reproduces exactly Eqs.~(\ref{continuity},\ref{density},\ref{2dimcurrent}). We have also checked that the same result is obtained by combining the time dependent equation with its adjoint using the explicit form of the Hamiltonian in Eq.~\eqref{hamiltonian1D}.   


We now proceed to compute the explicit form of the
current in terms of its azimuthal ($J_\varphi$) and radial ($J_r$) components $\bm{J}=J_\varphi \bm{e}_\varphi +J_r \bm{e}_r$, with  $\bm{e}_\varphi$ and $ \bm{e}_r$ being respectively the unit vectors of the azimuthal and radial directions: $\bm{e}_\varphi=-\sin\varphi\, \bm{e}_x+\cos\varphi\,\bm{e}_y$, and  $\bm{e}_r=\cos\varphi\, \bm{e}_x+\sin\varphi\,\bm{e}_y$. We also compute the divergence of the current $\nabla\cdot\bm{J}$ for the Dirac
fermion constrained to move on the ring of radius $R$ for which system the gradient operator is simply $\nabla=(\bm{e}_\varphi/R)\partial_\varphi$. A straightforward calculation shows that: 
\begin{subequations}
\label{current1}
\begin{align}
{\bm\nabla}\cdot\bm{J}&=\frac{1}{R}J_r+\frac{1}{R}\frac{\partial J_\varphi}{\partial\varphi}, \label{c1}\\
 J_r&=c(e^{-i\varphi}\chi_1^*\chi_2^{\phantom{*}}+c.c.),\label{c2}\\J_\varphi&=c(-ie^{-i\varphi}\chi_1^*\chi_2^{\phantom{*}}+c.c.).\label{c3}
\end{align}
\end{subequations}

The spectrum and corresponding eigenfunctions can be easily obtained and are given by (positive energy branch):
\begin{subequations}
\label{spectrum}
\begin{align}
\label{spectruma}
E_\ell &={\sqrt{m^2c^4+ \left(\frac{\hbar c}{R}\right)^2 (\ell-\beta) (\ell -\beta +1)}}\nonumber\\ & = \varepsilon_\ell \, m c^2 \qquad\qquad \ell=0,\pm 1,\pm 2 \cdots \\
\label{spectrumb}\varepsilon_\ell &=\sqrt{1+\left(\frac{\lambdabar_C}{R}\right)^2 (\ell-\beta) (\ell-\beta+1)} 
\end{align}
\end{subequations}
with $\lambdabar_C=\hbar/(mc)$ the reduced Compton wavelength of the par\-ti\-cle of mass $m$.
The eigenstates of the Hamiltonian are:
\begin{equation}
\label{Eq:eigenstates}
    \psi_\ell = {\cal N}_\ell \left(\begin{array}{c}  e^{i\ell \varphi}\\ \frac{\hbar c/R}{E_\ell +mc^2}\,
    i \left(\ell-\beta\right) \, e^{i(\ell+1)\varphi}\end{array}
    \right)
\end{equation}
with ${\cal N}_\ell$ a  normalization constant which, without loss of generality, can be taken to be real. It can be easily verified that the eigenstates are orthogonal and the normalisation constant ${\cal N}_\ell$ can be obtained  from  the standard  condition
\begin{equation}
\oint_{{\cal C}_R}dz~ \psi_{\ell^\prime}^\dag\psi_\ell^{\phantom{\dag}}=\int_0^{2\pi} R~ d\varphi\,  \psi_{\ell^\prime}^\dag(\varphi) \psi_\ell^{\phantom{\dag}}(\varphi)=\delta_{\ell^\prime \ell}\, ,
\end{equation}
${\cal C}_R$ being the ring of radius $R$. The result for the normalisation constant  ${\cal N}_\ell$ can be cast as:
\begin{subequations}
 \begin{align}
 {\cal N}_\ell &  = \frac{1}{\sqrt{R}}\, A_{\ell}\,; \\ 
 \label{coefficientsAL}
    A_\ell&=\frac{1}{\sqrt{2\pi}}     \left[ 1+{\left(\frac{\hbar c}{R}\right)^2 \frac{(\ell-\beta)^2}{(E_\ell+mc^2)^2}}\right]^{-1/2} \,.
 \end{align}
\end{subequations}
It may be pointed out that the eigenstates (\ref{Eq:eigenstates}) are also eigenstates of $J_z=L_z+\frac{\hbar}{2}\sigma_z$ with eigenvalues $\hbar (\ell+\frac{1}{2})$. Note that in the nonrelativistic limit ($c\to \infty$) 
 we have $E_\ell \to mc^2$ and $\lambdabar_{\text{C}}\to 0$ and $A_\ell \to 1/\sqrt{2\pi}$.
 On the other end in the ultra-relativistic limit  ($ E_\ell \gg mc^2)$ we obtain:
 \begin{equation}
 \label{Aellmassless}
     A_\ell \longrightarrow \frac{1}{\sqrt{2\pi}}\left(1+\frac{\ell-\beta}{\ell-\beta+1}\right)^{-1/2} \,.
 \end{equation}
The eigenstate $\psi_\ell$ in Eq.~\eqref{Eq:eigenstates} has a constant azimuthal current: 
\begin{equation}
\label{Eq:currentJell}
    J_\varphi^{(\ell)} = \frac{ A_\ell ^2}{E_\ell +mc^2} \, \frac{2\hbar c^2}{R^2} \, {\left(\ell-\beta\right)} \, .
    \end{equation}
Note that, denoting by $ \lceil x \rceil$ the Ceiling function of $x$, if $\ell \ge \lceil \beta \rceil $ then the azimuthal current is positive semi-definite, $  J_\varphi^{(\ell)} \ge 0$. The nonrelativistic limit of the azimuthal current in Eq.~\eqref{Eq:currentJell} reduces to 
\begin{equation}
\label{Eq:currentJellNR}
    J_\varphi^{(\ell)} \to \frac{\hbar}{2\pi m R^2}\, {\left(\ell-\beta\right)}
\end{equation}
which coincides exactly with the nonrelativistic current discussed in \cite{Goussev2020:ab}. On the other end it is worthwhile pointing out that for the eigenstate $\psi_\ell$ the corresponding radial current $J_r^{(\ell)}$ vanishes identically.

\section{Relativistic Quantum Backflow on a Ring}
\label{Sec:QBF}
In this section we discuss the quantum backflow of the relativistic fermion constrained on the ring of radius $R$. We follow closely the approach discussed in \cite{Goussev2020:ab} for the nonrelativistic particle highlighting differences and similarities.

We consider a generic physical state given by a linear combination of the energy eigenstates $\psi_\ell(\varphi,t)$ with $\ell\ge \lceil \beta \rceil$:
\begin{equation}
\label{Eq:state}
    \Psi(\varphi, t) = \sum_{\ell=\lceil \beta \rceil}^{+\infty}\, c_\ell \, \psi_\ell (\varphi,t)
\end{equation}
which we assume to be normalized, and thus the coefficients $c_\ell$ satisfy:
\begin{equation}
\label{Eq:constraint_cl}
    \sum_{\ell=\lceil \beta \rceil}^{+\infty}\, |c_\ell|^2 \, = 1\,.
\end{equation}
Before proceeding further, let us mention that as the particle is constrained to move on the circumference of a circle of fixed radius $R$, we shall consider only the azimuthal current $J_\varphi$ and we will investigate the azi\-muthal backflow in line with what was done in ref.~\cite{strange}. In this context it may also be noted that, although the radial current does not vanish  in general, the total radial flux across the ring can be shown to always be zero. This is easily understood since the radial current turns out to be a periodic function of $\varphi$, see the Appendix for details.

The azimuthal probability current $J_\varphi^{(\Psi)}(\varphi,t)$ corresponding to the state $\Psi$ in Eq.~\eqref{Eq:state} through Eq.~(\ref{c3}) reads:
\begin{eqnarray}
\label{Eq:current_psi}
J^{(\Psi)}_\varphi &=&  \frac{\hbar c^2}{R^2}\!\!\! \sum_{\ell,\ell'=\lceil \beta \rceil}^\infty c_\ell^* A^{\phantom{*}}_\ell\left[\frac{{{\ell-\beta}}}{E_\ell +mc^2}+\frac{{\ell'-\beta}}{E_{\ell'} +mc^2} \right] A_{\ell'} c_{\ell'}\nonumber\\
&&\times e^{\frac{i}{\hbar} (E_\ell -E_{\ell'})t} e^{-i(\ell -\ell')\varphi} \,.
\end{eqnarray}
For the state $\Psi$ the integrated probability flowing through a fixed point (say $\varphi=0$) in the azimuthal direction ($P_\Psi)$ in a time window $[-T/2,T/2]$ is :
\begin{equation}
\label{Eq:Ppsi}
    P_\Psi = \int_{-\frac{T}{2}}^{+\frac{T}{2}} \, dt\, J_\varphi^{(\Psi)}(0,t)\, ,
\end{equation}
and 
for the state in Eq.~\eqref{Eq:state} given the time dependence of the current in Eq.~\eqref{Eq:current_psi} we just need the  time  integral:
\begin{subequations}
\begin{align}
\int_{-\frac{T}{2}}^{+\frac{T}{2}} e^{\frac{i}{\hbar}\left(E_{\ell}-E_{\ell^{\prime}}\right) t} d t &= \left\{\begin{array}{ll}
T & \ell=\ell^{\prime} \\
\hbar \left. \displaystyle\frac{e^{{\frac{i}{\hbar}}\left(E_{\ell}-E_{\ell^\prime}\right) t}}{i\left(E_{\ell}-E_{\ell^\prime}\right)}\right|_{-\frac{T}{2}}^{+\frac{T}{2}} 
 & \ell \ne \ell^{\prime}
\end{array}\right. \nonumber\\
&= T\, \text{sinc} \left[\frac{ T\, \left(E_\ell -E_{\ell^\prime}\right)}{2 \hbar}\right] \nonumber
\end{align}
\end{subequations}
so that we finally obtain $P_\Psi$ in the form:
\begin{equation}
\label{Eq:P_psi2}
    P_\Psi = \sum_{\ell,\ell' =\lceil \beta \rceil}^{+\infty} c_\ell^*\, K_{\ell,\ell'}\, c_{\ell'}
\end{equation}
with
\begin{eqnarray}
\label{Eq:kernel1}
    K_{\ell,\ell'} &=& \frac{\hbar c^2 T}{R^2}\, A_\ell \, \left( \frac{{{\ell-\beta}}}{E_\ell+mc^2}+ \frac{{{\ell'-\beta}}}{E_{\ell'}+mc^2}\right)A_{\ell'}\times\nonumber\\
    &&\phantom{xxxx}\text{sinc}\left[\frac{T(E_\ell-E_{\ell'})}{2\hbar}\right] \,.
\end{eqnarray}
Defining the dimensionless quantity:
\begin{equation}
\label{alpha}
    \alpha= \frac{\hbar T}{4 m R^2}
\end{equation}
the kernel $K_{\ell,\ell^\prime}$ can be rewritten:
\begin{eqnarray}
\label{Eq:kernel2}
K_{\ell,\ell'} &=& \left(\frac{\alpha}{\pi}\right)\, 4\pi\, A_\ell \, \left( \frac{\ell-\beta}{1+\varepsilon_\ell}+ \frac{\ell'-\beta}{1+\varepsilon_{\ell'}}\right)A_{\ell'}\times\nonumber\\
    &&\phantom{xxxx}\text{sinc}\left[
    2 \alpha \left(\frac{R}{\lambdabar_{\text{C}}}\right)^2(\varepsilon_\ell -\varepsilon_{\ell^\prime})\right] \,.    \end{eqnarray}
We begin the study of the relativistic backflow on the ring by considering a quantum state with only two eigenstates, $\psi_{\ell_1}$ and $\psi_{\ell_2}$ with $\lceil\beta\rceil\le \ell_1 < \ell_2$.  Thus in Eq.~\eqref{Eq:state} we parametrize the coefficients $c_\ell$ as:
\begin{equation}
c_{\ell}=\left\{\begin{array}{ll}
\cos \frac{\delta}{2} &  \quad \ell=\ell_{1} \geq \lceil\beta\rceil \\
e^{i \gamma} \sin \frac{\delta}{2} & \quad \ell=\ell_{2}>\ell_{1} \\
0 & \quad \ell \ne \ell_1, \ell_2
\end{array}\right.
\label{Eq:parametrization}
\end{equation}
with $0\le \delta \le \pi $ and $0\le \gamma < 2\pi$. This choice of parameters
 will ensure that the normalization condition in Eq.~\eqref{Eq:constraint_cl} is satisfied. Inserting Eq.~\eqref{Eq:parametrization} into Eq.~\eqref{Eq:P_psi2} using either Eq.~\eqref{Eq:kernel1} or Eq.~\eqref{Eq:kernel2} we obtain:
\begin{equation}
P_\Psi^{(\ell_1,\ell_2)} = \frac{\alpha}{\pi} \Big[ A -B \cos\delta +C\, \text{sinc}(D) \cos\gamma\sin\delta\Big]
\end{equation}
where the quantities $A,B,C$ and $D$ are given by:
\begin{equation}
    \begin{array}{l}
    \displaystyle
    A= 4\pi\left(A_{\ell_2}^2\,\frac{ \ell_2 -\beta}{1+\varepsilon_{\ell_2}}+ A_{\ell_1}^2 \frac{\ell_1-\beta}{1+\varepsilon_{\ell_1}}\,   \right)  \\  \displaystyle  B= 4\pi\left(A_{\ell_2}^2\,\frac{ \ell_2 -\beta}{1+\varepsilon_{\ell_2}}- A_{\ell_1}^2 \frac{\ell_1-\beta}{1+\varepsilon_{\ell_1}}\,
    \right)\\
    \displaystyle
    C= 4\pi A_{\ell_2}A_{\ell_1}\left(\frac{\ell_2-\beta}{1+\varepsilon_{\ell_2}}+ \frac{\ell_1-\beta}{1+\varepsilon_{\ell_1}}\right)\\
    \displaystyle
    D= 2\alpha \left(\frac{R}{\lambdabar_{\text{C}}}\right)^2\left(\varepsilon_{\ell_2}-\varepsilon_{\ell_1}\right) \,.
    \end{array}
\end{equation}
We then perform a minimisation procedure over the parameters $\delta, \gamma$:
\begin{equation}
    {\cal P}^{(\ell_1,\ell_2)}  (\alpha,\beta) = \min\nolimits_{\left\{\delta,\gamma\right\}} P^{(\ell_1,\ell_2)}_\Psi (\delta,\gamma)
\end{equation}
and, similarly to what is found in \cite{Goussev2020:ab}, the result is:
\begin{equation}
\label{Pl1l2}
    {\cal P}^{(\ell_1,\ell_2)}  (\alpha,\beta) = \frac{\alpha}{\pi} \left(
    A -\sqrt{B^2 +C^2\, \text{sinc}^2 (D)}
    \right)\,.
\end{equation}
Note however that the coefficients $A, B, C$ and $D$ have a different definitions with respect to that of the  nonrelativistic problem.
\begin{figure}[h!]
    \includegraphics[width=0.95
    \linewidth]{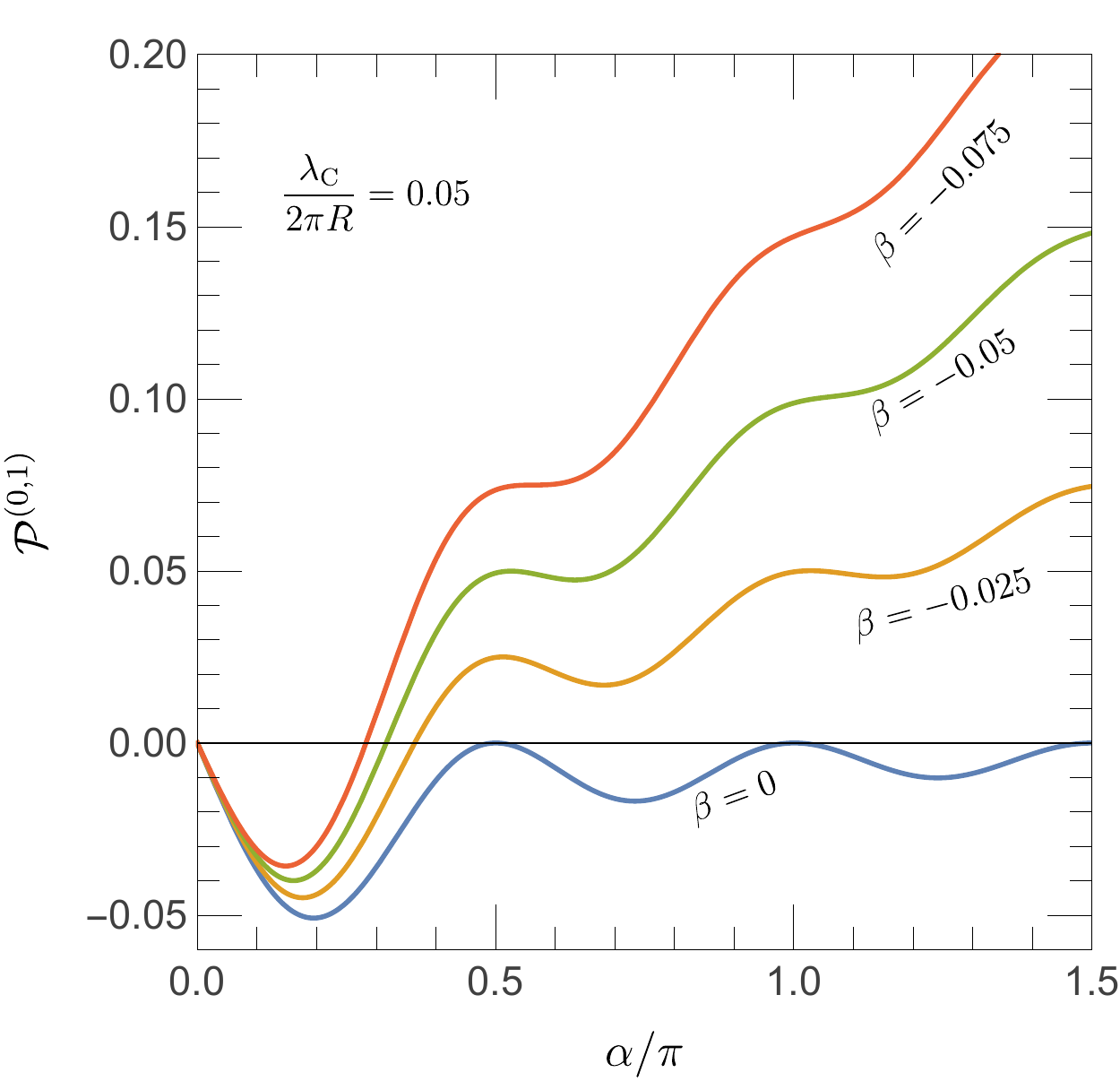}
    \caption{Plot of the backflow probability ${\cal P}^{(0,1)}$ at $\varphi=0$ for the relativistic fermion on the ring for  $\lambdabar_{\text{C}}/R =\lambda_{\text{C}}/(2\pi R)=0.05$. The solid (blue) line is the backflow probability ${\cal P}^{(0,1)}$ corresponding to a vanishing magnetic field ($\beta=0)$.  The orange, green and red  lines are the probabilities ${\cal P}^{(0,1)}$ for the relativistic problem corresponding respectively to $\beta= -0.025,-0.05,-0.075$.  }
    \label{fig:FIG1}
\end{figure}
In Fig.~\ref{fig:FIG1} we show the backflow probability ${\cal P}^{(0,1)}$ for different values of the parameter $\beta$ and for  $\lambdabar_{\text{C}}/R=0.05$.
Fig.~\ref{fig:FIG1} shows clearly that the probability can indeed become negative. We see that the lowest value, $\text{min}_{(\alpha,\beta)} {\cal P}^{(0,1)} \approx -0.0508602  $ is attained for $\beta=0$ at $\alpha/\pi\approx 0.195067$.

\subsection{ General Formulation}

Here we shall consider a more general superposition of eigenstates, and study the  corresponding problem of maximizing the integrated probability flux through $\varphi=0$, given in Eq.~\eqref{Eq:P_psi2}, 
i.e. maximize the quantum backflow considering a generic state as in Eq.~\eqref{Eq:state} subject to the normalization constraint Eq.~\eqref{Eq:constraint_cl}.
A comment is in order here about a symmetry of our system. Indeed it can easily be checked that the eigenvalues, cf. Eq.~\eqref{spectruma}, satisfy the following relation:
\begin{equation}
E_\ell^{(\beta)} = E_{\ell-1}^{(\beta-1)}
\end{equation} and the same is true for the normalization constants, Eq.~\eqref{coefficientsAL}. This in turn implies that the  quantity ${\cal P}_\psi$ as given in Eq.~\eqref{Eq:P_psi2} is invariant under the transformation $\beta \to \beta-1$ and $c_\ell \to c_{\ell+1} $ (or equivalently $\beta \to \beta+1$ and $c_\ell \to c_{\ell-1} $). Therefore, it will be enough to consider $\beta $ in the interval:
\begin{equation}
\beta \in (-1, 0] \qquad \text{with} \qquad \lceil\beta\rceil=0\,.
\end{equation}
Hereafter we shall therefore assume $\lceil\beta\rceil=0$.

As  also discussed in \cite{Goussev2020:ab} the maximal backflow   can be found  defining   a   real-valued functional of the set of coefficients $\left\{ c_\ell \right\}$ by:
\begin{equation}
I\left[c_{\ell}\right]=\sum_{\ell, \ell'=0}^{\infty} c_{\ell}^{*} K_{\ell \ell'} c_{\ell'}-\lambda \sum_{\ell=0}^{\infty} c_{\ell}^{*} c_{\ell}
\end{equation}
and performing an unconstrained minimization
where $\lambda$ plays the role of a Lagrange multiplier. The variation of the functional $I[\left\{ c_\ell \right\}]$ with respect to independent variations $\delta c_\ell$ and $\delta c_\ell^*$ produces the corresponding eigenvalue problem (Euler-Lagrange equations):
\begin{equation}
\label{Eq:eigenvalue}
\sum_{\ell'=0}^{\infty} K_{\ell \ell'} c_{\ell'}=\lambda\, c_{\ell}
\end{equation}
so that the maximal amount of quantum backflow is attained when the value of $P_\Psi$,  Eq.~\eqref{Eq:P_psi2}, reaches its infimum and this coincides with the lowest eigenvalue of the spectrum \{$\lambda$\} of the  unconstrained variational problem:
\begin{equation}
\label{Eq:infimum}
\mathcal{P}(\alpha, \beta,\frac{\lambdabar_{\text{C}}}{R}) \equiv \inf _{\Psi} P_{\Psi}=\inf \{\lambda\}\, .
\end{equation}
We note that the minimisation procedure leading to Eq.~\eqref{Eq:eigenvalue} and Eq.~\eqref{Eq:infimum}, i.e. that the maximal amount of backflow is related to the lowest (negative) eigenvalue of the kernel in Eq.~\eqref{Eq:kernel2} is a simple application of the min-max  (Rayleigh quotient) theorem~\cite{gelfand2012calculus}. 

We conclude this subsection with a comment regarding the fact that the presence of a nonvanishing radial current does not affect at all the analysis of the azimuthal backflow presented here. First of all, we note that since the kernel $K_{\ell,\ell'}$ is real and symmetric the eigenvectors of the eigenvalue problem in Eq.~\eqref{Eq:eigenvalue} are real. This means that  the state $\Psi$ maximising the azimuthal backflow is given by Eq.~\eqref{Eq:state} with all coefficients $c_\ell$ real. 
For a state $\Psi$ built with real coefficients  the integral of the  radial current at $\varphi=0$, $J_r(0,t)$, over a symmetric time interval  always vanishes, see Eq.~\eqref{Eq:Prpsi}, whereas the corresponding time integrated azimuthal current  (cf. $P_\Psi$ in Eq.~\eqref{Eq:state}) reaches its infimum value (its absolute value is maximised). Therefore any worry that a nonzero radial component of the current might affect the azimuthal backflow analysis presented here should be discarded at once. 
At last, we also note that  $\varphi=0$ is just one point on the ring and therefore given the axial symmetry of the problem at hand, the above conclusion will hold as well at any other point. 

In the next subsection we will discuss the numerical solution of the eigenvalue equation in Eq.~\eqref{Eq:eigenvalue}, then in the next section we will address  the line limit of backflow on a ring which is obviously reached when $R\to \infty$.
\begin{figure*}[t]
    \includegraphics[width=0.475
    \linewidth]{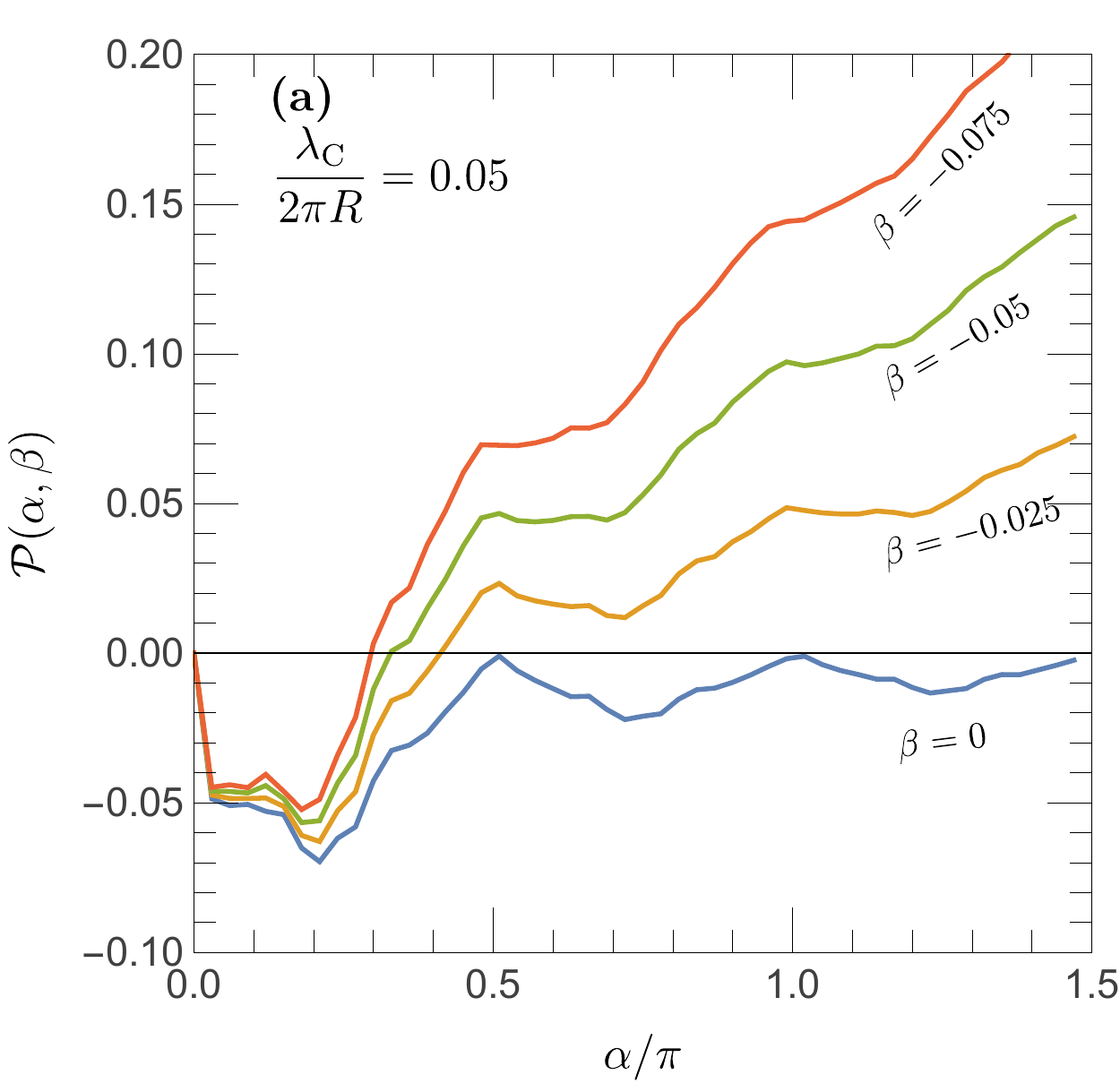}\hspace{0.3cm}
    \includegraphics[width=0.475
    \linewidth]{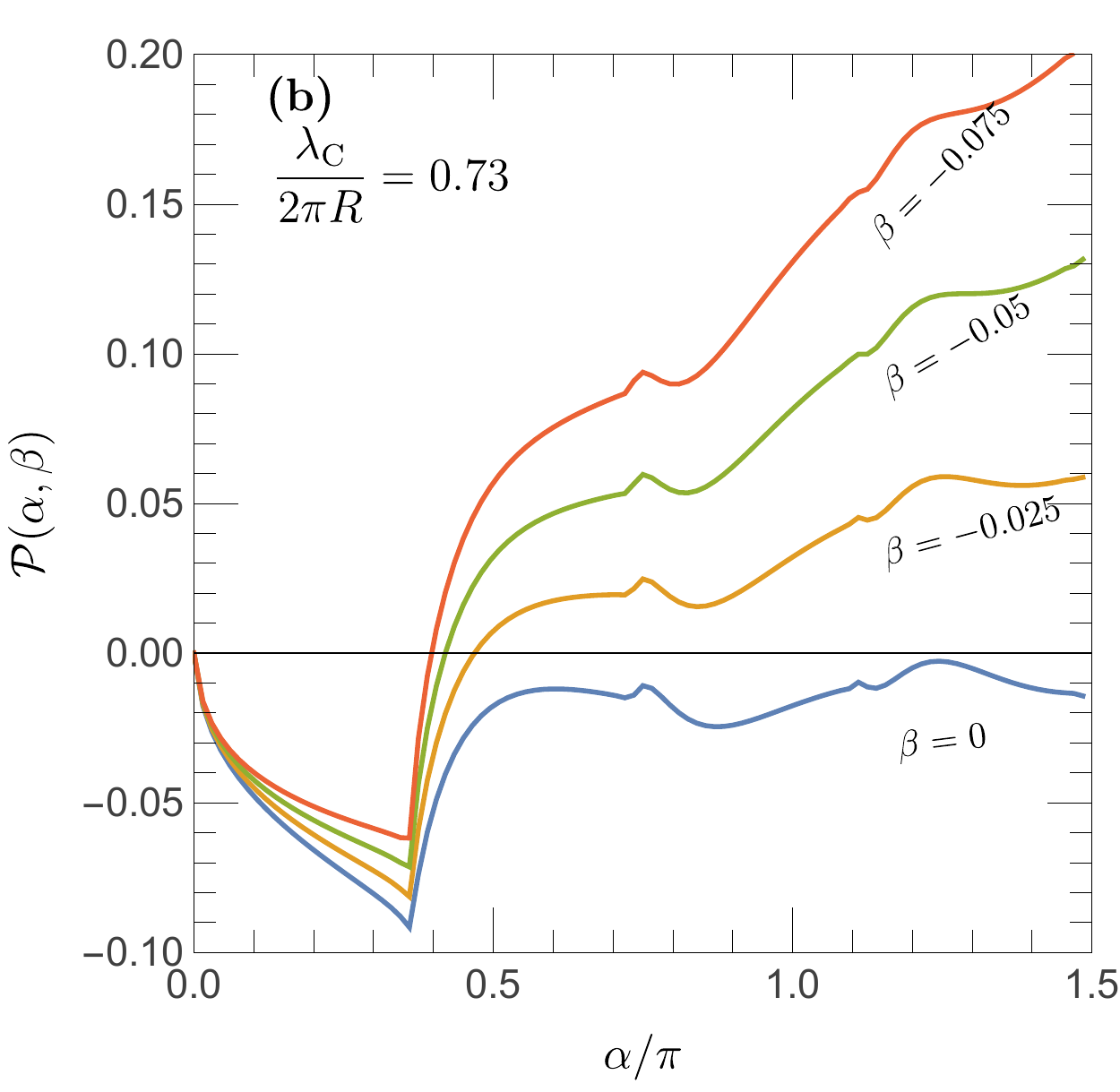}
    {\includegraphics[width=0.475\linewidth]{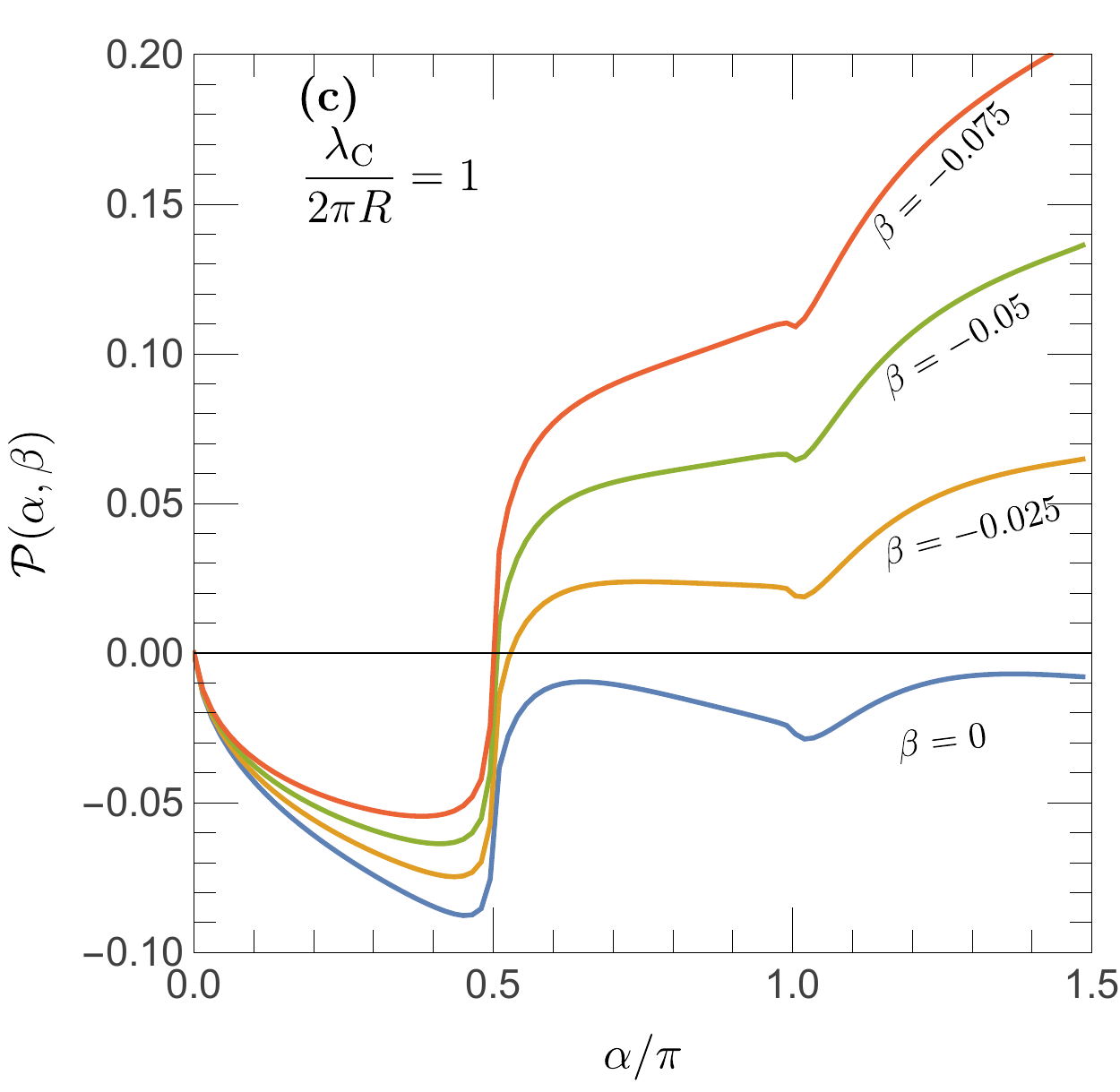}}\hspace{0.3cm}
    \includegraphics[width=0.475
    \linewidth]{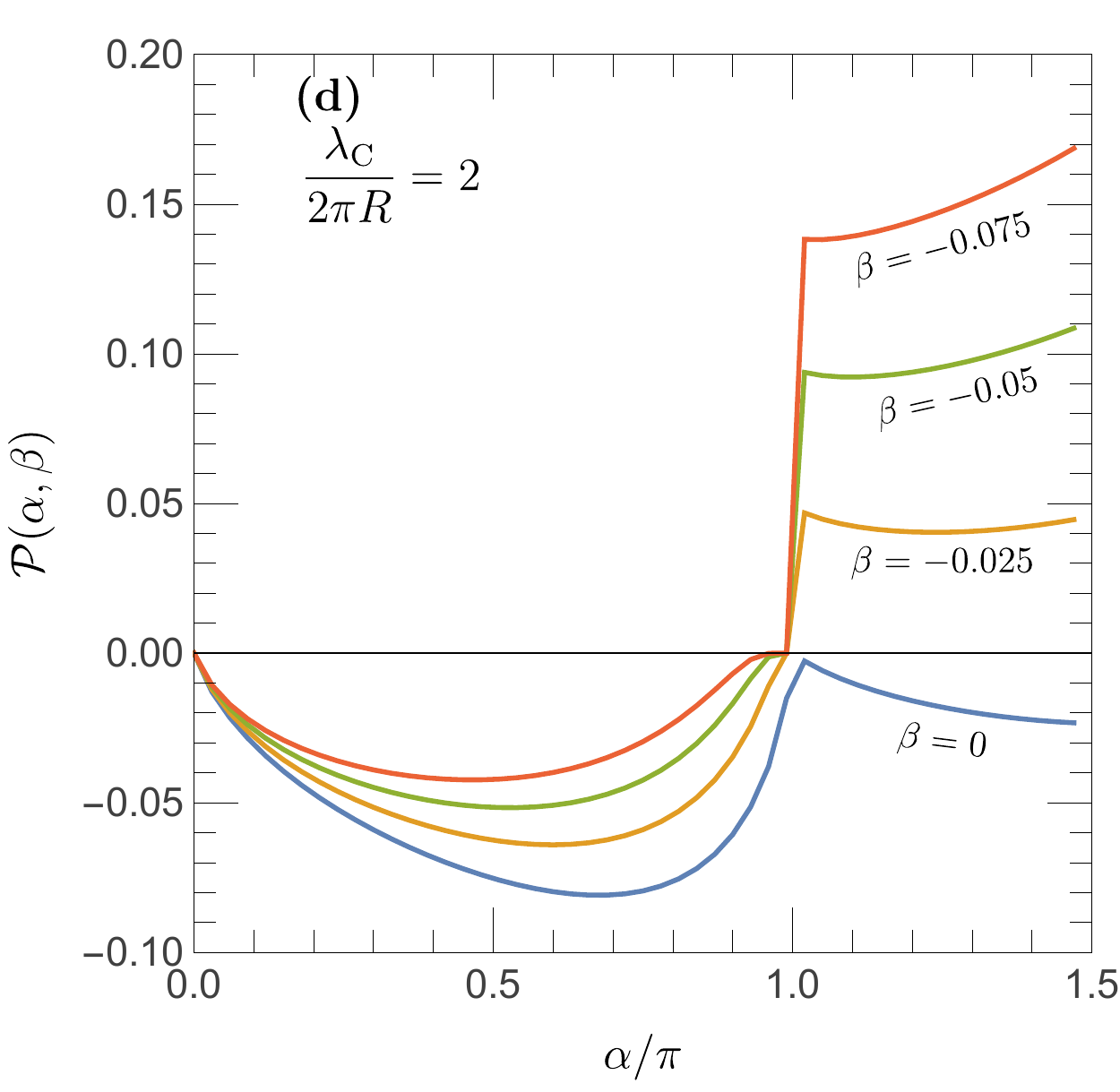}
    \caption{$\mathcal{P}(\alpha,\beta)$ for a fermion of mass $m$, on the ring of radius $R$ for various values of the parameter $\beta$ and for different values of $\lambdabar_{\text{C}}/R=\lambda_{\text{C}}/(2\pi R)$: top left panel (a) $\lambdabar_{\text{C}}/R=0.05$, top right panel (b) $\lambdabar_{\text{C}}/R=0.73$, bottom left panel (c) $\lambdabar_{\text{C}}/R=1$ , bottom right panel (d) $\lambdabar_{\text{C}}/R=2$. We observe a quite different structure depending on the value of the parameter $\lambdabar_{\text{C}}/R$. }
    \label{fig:FIG2}
\end{figure*}
\begin{figure}[htb!]
    \includegraphics[width=0.95
    \linewidth]{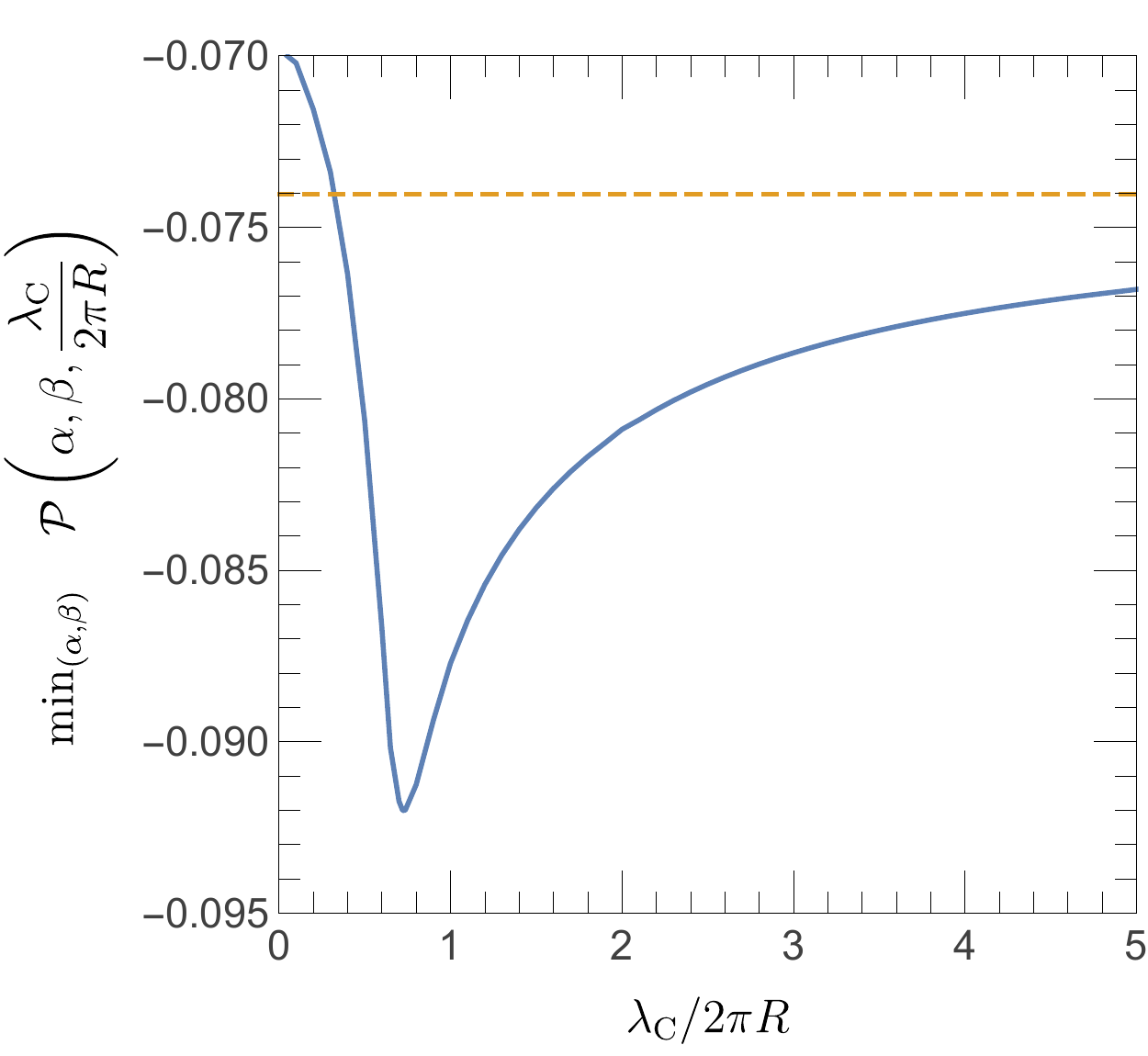}
\caption{\label{FIG:3} (Color online) Numerical results of the minimum eigenavule of the Kernel as a function of the parameter $\lambdabar_{\text{C}}/R=\lambda_{\text{C}}/(2\pi R)$. We can appreciate an important structure with respect to this parameter. We observe a maximal amount of backflow for $\lambdabar_{\text{C}}/R=0.730$. The smallest value of the parameter $\lambdabar_{\text{C}}/R$  investigated in the plot is $\lambda_{\text{C}}/(2\pi R)=0.05$. See Table.~\ref{Table:II} for smaller values of $\lambda_{\text{C}}/(2\pi R)$. } The dashed line (orange) represent the preliminary result for the limiting value of the backflow as $\lambdabar_{\text{C}}/R\to \infty$ as we have checked values  from 20 up 100.000 that are shown in Table~\ref{Table:I}. 
\end{figure}

\subsection{Numerical analysis}
\label{Sec:numerical}
Following the procedure described in \cite{Goussev2020:ab}, we truncate the sum in Eq.~\eqref{Eq:eigenvalue} at a large value $\ell'=N$, and find the minimum eigenvalue $\lambda^{(N)}_{\text{min}}$ of this finite-dimensional problem. By repeating this for a sequence of increasing $N$ we obtain a numerical estimate for $\mathcal{P}(\alpha, \beta)=\lim_{N\to\infty}\lambda^{(N)}_{\text{min}}$ by plotting the data as a function of $1/N$ with a quadratic polynomial and extracting the value at $1/N=0$. The choice of a quadratic polynomial is justified by the values of the sum of squared residuals, with orders generally less than or equal to $10^{-10}$. $\mathcal{P}(\alpha, \beta)$ is also a function of the parameter $\lambdabar_{\text{C}}/R$, which is shown in Fig.~\ref{fig:FIG2}, by plotting it as a function of $\alpha$ for different values of $\beta$ and $\lambdabar_{\text{C}}/R$. A detailed numerical analysis reveals that the absolute maximal amount  of  backflow is found for $\lambdabar_{\text{C}}/R=0.730$:
\begin{equation}
\label{Eq:mainresult}
    \left.\inf_{\alpha,\beta}\mathcal{P}\right|_{\lambdabar_{\text{C}}/R = 0.730}=-c_{\text{Dring}},\qquad c_{\text{Dring}}\simeq0.091999
\end{equation}
 with the minimum attained for $\beta=0$ and $\alpha/\pi\simeq0.36252$. 
\begin{table*}[hbt!]
\caption{\label{Table:I}Numerical estimate of the  quantum backflow for large values of the parameter $\lambdabar_{\text{C}}/R$. This corresponds to estimating the QBF for a massless particle since  $\lambdabar_{\text{C}}/R \to \infty$ is equivalent to the massless limit  ($m\to 0$).}
\begin{ruledtabular}
\begin{tabular}{cccccc}
$\frac{\lambdabar_{\text{C}}}{R} $    & 20 &500&1000&10000&100000 \\
$\text{inf}_{\psi} {\cal P}_\psi$     & -0.074726&-0.074061& -0.074047& -0.074034&-0.074034
\end{tabular}
\end{ruledtabular}
\end{table*}
\begin{table*}[hbt!]
\caption{\label{Table:II}Numerical estimate of the quantum backflow for small values of the parameter $\lambdabar_{\text{C}}/R$. This corresponds to estimating the QBF in the non relativistic limit since   $\lambdabar_{\text{C}}/R \to 0$ is equivalent to   ($c\to \infty$).}
\begin{ruledtabular}
\begin{tabular}{cccccc}
$\frac{\lambdabar_{\text{C}}}{R} $    & 0.05 &0.01&0.001&0.0001&0.00001 \\
$\text{inf}_{\psi} {\cal P}_\psi$     & -0.070001& -0.069560& -0.069416&-0.069417&-0.069418
\end{tabular}
\end{ruledtabular}
\end{table*}
This was obtained by evaluating the infimum separately for different values of $\lambdabar_{\text{C}}/R$ which results in Fig.~\ref{FIG:3}.

We have estimated the large $\lambdabar_{\text{C}}/R\to \infty$ limit (which corresponds to the massless fermion case ($m\to 0$). While further investigation of the massless case will be the object of a future work, the preliminary results of these estimates are shown in Table~\ref{Table:I} where we can see that the value of the quantum backflow approaches 0.074 for large $\lambdabar_{\text{C}}/R$. This limiting value is shown in Fig.~\ref{FIG:3} as a horizontal (dashed) line.

Using the complete numerical results of the eigensystem in Eq.~\eqref{Eq:eigenvalue}, i.e. the eigenvector corresponding to the most negative  eigenvalue, we can find a numerical approximation to the current corresponding to the backflow-maximizing state for this system. We set $\alpha$, $\beta$ and $\lambdabar_{\text{C}}/R$ to the values corresponding to $\mathcal{P}\simeq-c_{\text{Dring}}$, truncate the sum in Eq.~(\ref{Eq:eigenvalue}) at $N=2000$, and compute the eigenvector corresponding to the smallest eigenvalue. This will provide the approximate coefficients $c_\ell$ up to $\ell_{\text{max}}=N=2000$ of the backflow-maximizing state, as defined in Eq.~(\ref{Eq:state}).

We shall now examine the backflow in the nonrelativistic regime i.e., ($c\to \infty$) and $\lambdabar_{\text{C}}/R \to 0$. To do this let us first note that in the nonrelativistic limit the Hamiltonian, spectrum and kernel are given by
\begin{subequations}   
\label{nr1}
\begin{align}
H_{\text{NR}} &=\frac{\hbar^2}{2m R^2}\left[(\hat{\ell}_z-\beta)^2 + \hat{\ell}_z-\beta\right]\label{nr1a}\\
(E_\ell)_{\text{NR}}&= mc^2 + \frac{\hbar^2}{2mR^2} \left[(\ell-\beta)^2 +\ell -\beta\right]\label{nr1b}
\end{align}
\end{subequations}
where $\hat{\ell}_z= -i\partial_\varphi$, $\ell=0,\pm 1, \pm 2\, ...\, ... $ and 
\begin{equation}
\label{Eq:NRkernel1}
    (K_{\ell,\ell'})_{\text{NR}} = \frac{\alpha}{\pi}(\ell +\ell' -2\beta)\, \text{Sinc}\left[ \alpha(\ell-\ell') (\ell+\ell'-2\beta+1)\right]\, .
\end{equation}
Now from Fig.~\ref{FIG:3} we see that our system in the non relativistic limit exhibits a maximal quantum backflow $(c_\text{Dring})_\text{non rel.}\approx 0.07$ for $\lambdabar_{\text{C}}/R=0.05 $ (the smallest value represented in Fig.~\ref{FIG:3}).
To see the origin of this value we have  computed the maximal backflow for several  values of $\lambdabar_{\text{C}}/R$ smaller than 0.05 and the results are shown in Table~\ref{Table:II}. These numerical findings  indicate that the non relativistic limit of the maximal backflow on the ring is $(c_\text{Dring})_\text{non rel.}\approx 0.069$. The actual non-relativistic limit of quantum backflow  on the ring ($\lambdabar_{\text{C}}/R \to 0$) can however also be obtained by studying the eigenvalue problem leading to the maximal quantum backflow, c.f Eq.~\eqref{Eq:eigenvalue}, but using the reduced kernel in Eq.~\eqref{Eq:NRkernel1} which is itself derived by using the non-relativistic reduced spectrum in Eq.~\eqref{nr1b}. The result of this modified eigenvalue  study is:
\begin{equation}
(c_\text{Dring})_\text{non rel.} =0.069418\,.    
\end{equation}
perfectly in line with expectations from what is suggested by the numerical study of computing the maximal backflow  at increasingly smaller values of $\lambdabar_{\text{C}}/R$ reported in Table~\ref{Table:II}.
Note that for the same quantity in a study~\cite{Goussev2020:ab} of a non relativistic particle on a ring of radius $R$, a value of  $c_\text{ring}\approx 0.116 816$ is found. The difference between these two values can be understood in terms of the fact that the Hamiltonian and  the spectrum  in Eq.~(\ref{nr1}), and consequently the kernel in Eq.~(\ref{Eq:NRkernel1}), are different from the corresponding quantities in the nonrelativistic model \cite{Goussev2020:ab}.

\begin{figure*}[htb]
    \includegraphics[scale=0.68]{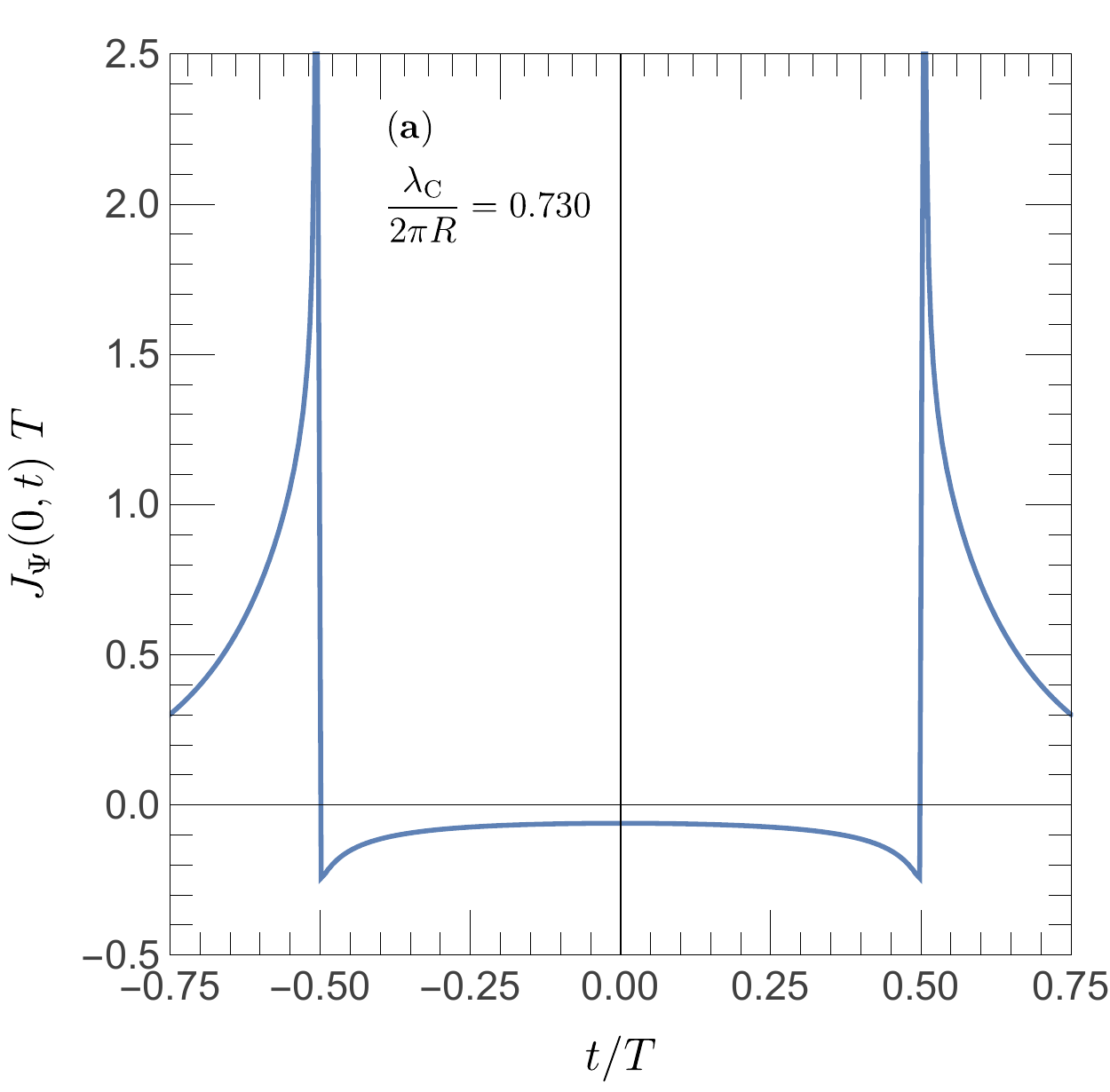}
    \includegraphics[scale=0.69]{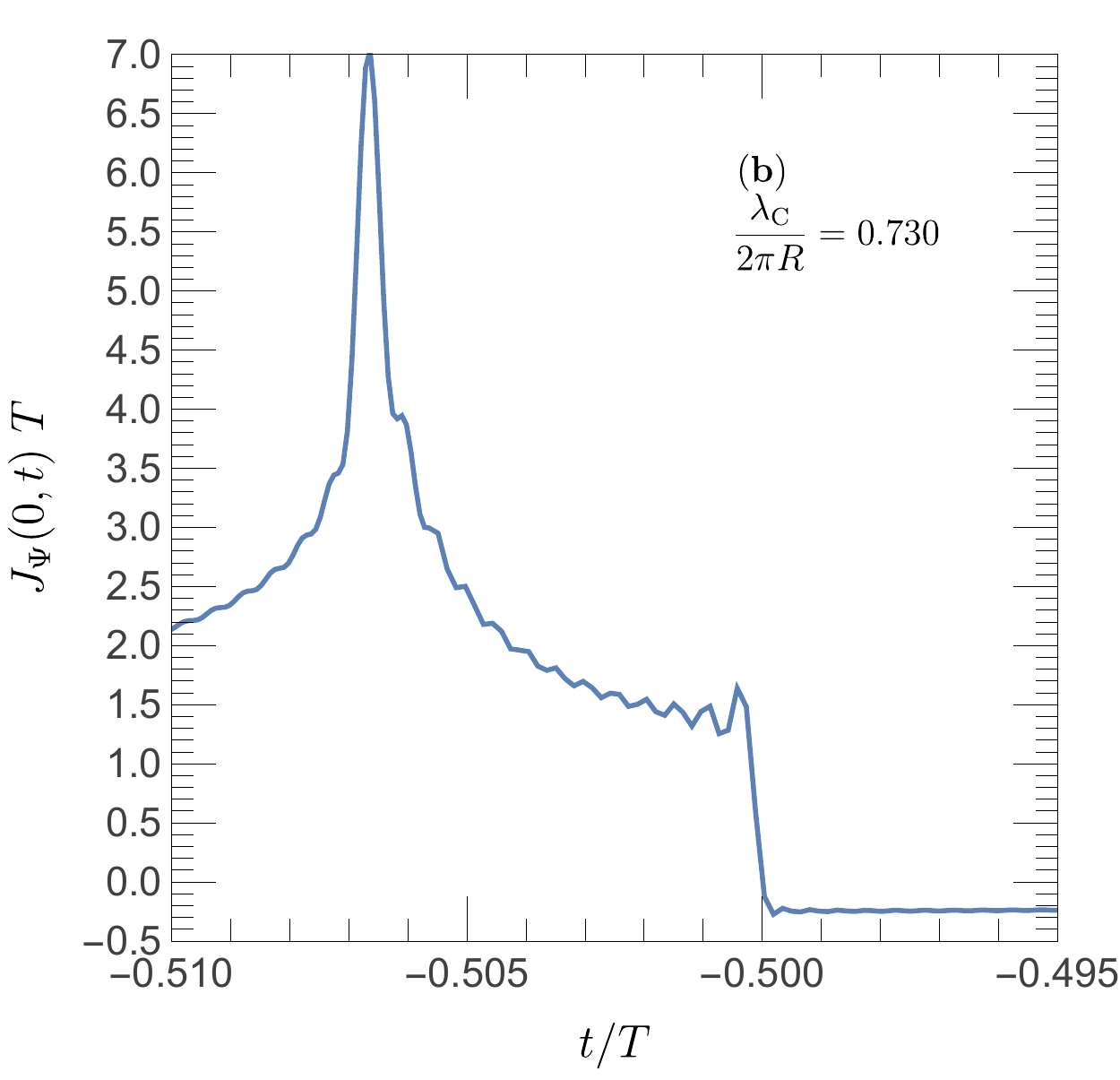}
\caption{\label{FIG:4} (a) Probability current $J_\Psi(0,t)T$ as a function of time $t$, in units of $1/T$, for the backflow-maximizing state. (b) We show a zoom in  the neighborhood of $t/T=-0.5$ where the current changes sign.  }
\end{figure*}

Finally we can  compute the corresponding probability current, Eq.~(\ref{Eq:current_psi}), at $\varphi=0$ as a function of time. It turns out that it is convenient to put the azimuthal probability current in the form: 

\begin{widetext}

\begin{equation}
J_\varphi^{(\Psi)}(0,t)\,T= 4\alpha\!\!\! \sum_{\ell,\ell'=0}^N c_\ell^* A_\ell\left[ \frac{\ell}{1+\varepsilon_\ell}+ \frac{\ell'}{1+\varepsilon_{\ell'}}\right] A_{\ell'} c_{\ell'}\,\, e^{i \left(\sqrt{\left(\frac{R}{\lambdabar_{\text{C}}}\right)^2+\ell(\ell+1)}-\sqrt{\left(\frac{R}{\lambdabar_{\text{C}}}\right)^2+\ell'(\ell'+1)}\right)\frac{4\alpha R}{\lambdabar_{\text{C}}}\frac{t}{T}}\\
\end{equation}

\end{widetext}
so that it is written in terms of quantities whose values are known for the backflow-maximizing state. The result is shown in Fig.~\ref{FIG:4}a
where it is clear that in the interval $t\in [-T/2, +T/2]$ the azimuthal current $J_\varphi^{(\Psi)} (0,t)$ corresponding to the maximizing state is always negative. Fig.~\ref{FIG:4}b shows an enlargement of the neighborhood $t/T\approx-1/2$ obtained by computing the current. While from Fig.~\ref{FIG:4}a one might think that at $t=-T/2$ there is a discontinuity the zoom  proposed in Fig.~\ref{FIG:4}b shows that in reality the current changes without discontinuity. This behaviour of the current is similar to what has been found in other one dimensional systems on the line \cite{Ashfaque:2019aa} as well as in the nonrelativistic problem of a particle on a ring~\cite{Goussev2020:ab}.   

\section{Line limit}
We discuss here the line limit of the quantum backflow on the ring. Let us consider the limit $R\to\infty$ of the eigenvalue equation, Eq.~\eqref{Eq:eigenvalue},  with $K_{\ell,\ell'}$ given in Eq.~\eqref{Eq:kernel2}. We note that: 
\begin{equation}
\varepsilon_\ell =\sqrt{1+\frac{\hbar^2}{m^2c^2R^2} \frac{1}{\alpha}\,\sqrt{\alpha}(\ell -\beta)\sqrt{\alpha}(\ell -\beta+1)
}\,.
\end{equation}

The line limit is reached when $\alpha\to 0$ with $\sqrt{\alpha}\beta \to 0$ so that using $\alpha= \hbar T /(4 m R^2)$ we have: 
\begin{equation}
\varepsilon_\ell \to \sqrt{1+\frac{4\hbar T^{-1}}{mc^2}\, (\sqrt{\alpha}\ell)^2}   \end{equation}
and  with the definitions:
\begin{equation}
\label{epsilon}
  \epsilon= \sqrt{\frac{4\hbar T^{-1}}{mc^2}}\,,  \qquad z=\sqrt{\alpha}\, \ell
\end{equation} 
the line limit of the quantity $\varepsilon_\ell$ is:
\begin{equation}
\label{gamma}
    \varepsilon_{\ell}\to \sqrt{1+\epsilon^2 z^2} \equiv \gamma(z)\,.
\end{equation}
A little algebra shows, with the same reasoning, that the coefficients $A_\ell$ in eq.~\eqref{coefficientsAL} reduce to:
\begin{equation}
\label{coefficientALz}
 A_\ell \to\frac{1}{\sqrt{2\pi}} \sqrt{\frac{\gamma(z)+1}{2\gamma(z)}} \,.  
\end{equation}
With this in mind,
the eigenvalue equation, Eq.~\eqref{Eq:eigenvalue}, can be first manipulated by  distributing $\sqrt{\alpha}$ in the numerators of the two terms in the round brackets in Eq.~\eqref{Eq:kernel2}. We thus find:
\begin{eqnarray}
&&\frac{1}{\pi}\,\sum_{\ell'=0}^{\infty} \, 4\pi\, A_\ell \, \left[ \frac{\sqrt{\alpha}(\ell-\beta)}{1+\varepsilon_\ell}+ \frac{\sqrt{\alpha}(\ell'-\beta)}{1+\varepsilon_{\ell'}}\right]A_{\ell'}\nonumber\times \\&&\phantom{xxxxxxxxxxx} \text{Sinc}\left[
    \frac{2}{\epsilon^2}(\varepsilon_\ell -\varepsilon_{\ell^\prime})\right] c_{\ell'} = \lambda \frac{c_\ell}{\sqrt{\alpha}}   \end{eqnarray}
  so that taking the limit $\alpha\to 0$ by defining the variables $z=\sqrt{\alpha}\,\ell$, $z'=\sqrt{\alpha}\,\ell'$, the function $f(z)=c_\ell/\sqrt{\alpha}$, using the results in Eqs.~(\ref{epsilon},\ref{gamma},\ref{coefficientALz})  and replacing the discrete sum over $\ell'$ by an integral over the variable $z'$ one finds:  
\begin{eqnarray}
&&\frac{1}{\pi}\!\int_{0}^{\infty} \,\!\!\! dz'\, \sqrt{\frac{\gamma(z)+1}{\gamma(z)}} \, \left[ \frac{z}{1+\gamma(z)}+ \frac{z'}{1+\gamma(z')}\right]\sqrt{\frac{\gamma(z')+1}{\gamma(z')}}\nonumber\\&&\phantom{xxx}\times  \text{sinc}\left[
    \frac{2}{\epsilon^2}(\gamma(z) -\gamma(z'))\right]\, f(z') = \lambda\, f(z)   \end{eqnarray}
that can be  straightforwardly rearranged into
\begin{eqnarray}
&&\frac{1}{\pi}\!\int_{0}^{\infty} \,\!\!\! dz'\, 
\frac{z(\gamma(z')+1)+z'(\gamma(z)+1)}{\sqrt{\gamma(z)(\gamma(z)+1)\,\gamma(z')(\gamma(z')+1)}}
\phantom{xxxxxxxx}\nonumber\\&&\phantom{xxxxx}\times  \text{sinc}\left[
    \frac{2}{\epsilon^2}(\gamma(z) -\gamma(z'))\right]\, f(z') = \lambda\, f(z) 
\label{Eq:massive_fermion_line}    
\end{eqnarray}
which coincides exactly with the integral eigenvalue equation for a relativistic spin-half fermion on the line discussed in \cite{Melloy1998:aa,Ashfaque:2019aa} (cf. see their equation 14).

\section{Discussion and Conclusions}
\label{Sec:conclusions}
In this paper we have presented a study of the quantum backflow of a charged relativistic massive fermion in a plane within a perpendicular, constant magnetic field and  constrained to move on a  ring of radius $R$. The same problem for a charged nonrelativistic spinless particle has also been studied \cite{Goussev2020:ab}. We use the same approach to the problem presented in \cite{Goussev2020:ab}  highlighting differences and similarities with the results of the nonrelativistic analysis. In this context it may be noted that an  experimental realization of the setup could be possible in a ring of a two dimensional material like gapped graphene where the quasiparticles are massive \cite{PhysRevB.78.085101,Belouad_2018}.

Before recalling our main results we note that constraining the particle in 2+1 dimensions  into a ring of radius $R$ reduces the system to one which is effectively 1+1 dimensional. If we were to start from a 3+1 dimensional Dirac equation and fix the radial coordinate at $r=R$ we would make the system effectively 2+1 dimensional (a relativistic particle moving on a sphere of radius $R$). This would somehow parallel previous studies,~\cite{strange,PhysRevA.102.062218}, (where a non-relativistic particle moves on a plane) where some issues were discussed such as that of a proper definition of the QB probability~\cite{PhysRevA.102.062218}. The treatment of the relativistic Dirac particle on a fixed radius sphere (again a two-dimensional system) could perhaps be addressed as well but it certainly goes beyond the scope of the present work.

The main finding reported here is that  the maximal amount of backflow for the relativistic Dirac fermion on the ring is obtained when the radius of the ring is of the order of the reduced Compton wavelength of the particle,  $\lambdabar_{\text{C}}/R=0.730$ (with $\beta=0$ and $\alpha/\pi\simeq0.36252$) and it amounts to  $c_{\text{Dring}}=0.092$. We have then  discussed the line limit showing that the eigenvalue problem reduces to the integral equation for the quantum backflow of a relativistic particle on a line previously studied in the literature~\cite{Ashfaque:2019aa}. 

We have also computed  the probability current flowing through the point $\varphi=0$ as a function of time for the state that maximises the net flow in the time interval [$-T/2,+T/2$], finding, as expected, that it is always negative in this interval.

It is also interesting to compare the result in Eq.~\eqref{Eq:mainresult} with the corresponding result obtained in \cite{Goussev2020:ab} for the nonrelativistic problem on the ring $c_{\text{ring non-rel}} = 0.116 816$.  We note that the amount of backflow for the Dirac fermion on a ring is about 20\% smaller than that of a nonrelativistic particle but still larger than the backflow value on the line $c_{\text{line}}=0.0389$. 

It is interesting to find out how our results look in the nonrelativistic limit. To this end we refer the reader to Eqns.\eqref{nr1}-\eqref{Eq:NRkernel1} which were used to compute the backflow. The results of this computation differ from those of \cite{Goussev2020:ab} and this difference is due to the fact that the nonrelativistic limit of a constrained spin $1/2$ Dirac system does not necessarily reduce to the Schr\"odinger equation for a spin $0$ particle.

Finally, we provide preliminary estimates of the amount of backflow in the limit $\lambdabar_{\text{C}}/R \to \infty$ which corresponds to the massless fermion case ($m\to 0)$. While we intend to study the massless case comprehensively in a separate work, these preliminary results  in our opinion show that there is definitely quantum backflow for a massless Dirac fermion on a ring.

\appendix*
\section{Radial current}
\begin{widetext}

The radial probability current $J_r^{(\Psi)}$ as in Eq.~\eqref{c2} corresponding to the general state {$\Psi$} in Eq.~\eqref{Eq:state}, is
\begin{equation}
J_r^{(\Psi)}(\varphi,t)=  \frac{\hbar c^2}{R^2}\!\!\! \sum_{\ell,\ell'=\lceil \beta \rceil}^\infty \operatorname{Re}\,i\,c_{\ell'}^* A_{\ell'}\left[\frac{{{\ell-\beta}}}{E_\ell +mc^2}-\frac{{\ell'-\beta}}{E_{\ell'} +mc^2} \right] A_{\ell} c_{\ell}
\,\, e^{-\frac{i}{\hbar} (E_\ell -E_{\ell'})t} \,\,e^{i(\ell -\ell')\varphi} \,.
\end{equation}
The probability flux over the ring ${\cal C}_R$ along the radial direction is given by:
\begin{equation}
 \oint_{{\cal C}_R}dz~ \bm{J}^{(\Psi)}\cdot \bm{e}_r =\int_0^{2\pi} R~ d\varphi\, J_r^{(\Psi)} (\varphi,t)\,.   
\end{equation}
Integrating $\varphi$ over the interval $\left[0,2\pi\right]$ results in
\begin{equation}
\oint_{{\cal C}_R}dz~ \bm{J}^{(\Psi)}\cdot \bm{e}_r =  2\pi \frac{\hbar c^2}{R}\!\!\! \sum_{\ell,\ell'=\lceil \beta \rceil}^\infty\!\! \operatorname{Re}\, i\, c_{\ell'}^* A_{\ell'}\left[\frac{{{\ell-\beta}}}{E_\ell +mc^2}-\frac{{\ell'-\beta}}{E_{\ell'} +mc^2} \right] A_{\ell} \,c_{\ell}
\,\, e^{-\frac{i}{\hbar} (E_\ell -E_{\ell'})t} \, e^{i(\ell -\ell')\pi} \,\text{Sinc}\left[\left(\ell-\ell'\right)\pi\right] \,.
\label{Eq:flux}
\end{equation}
In the expansion in Eq.~\eqref{Eq:flux} each term is equal to $0$ both for $\ell=\ell'$ and $\ell\neq\ell'$, which means that the total radial probability flux over the ring always vanishes for the most general state $\Psi$ given by Eq.~(\ref{Eq:state}).
It is interesting to note that in a non-relativistic setting results similar to ours can be observed. For example, it can be shown that in the framework of ref \cite{strange} the radial current for an individual eigenstate vanishes while it is non vanishing for a general superposition state and the total radial flux is zero.
It can also be easily shown that if the state $\Psi$ given by Eq.~(\ref{Eq:state}) is chosen with general complex coefficients $c_\ell =|c_\ell| e^{i\gamma_\ell} $ than the integral of $J^{\Psi}_r(0,t)$ over a symmetric time integral can be cast as:
\begin{equation}
    \int_{-T/2}^{+T/2} dt J^{\Psi}_r(0, t) = - \frac{\hbar c^2}{R^2} T \!\!\! \sum_{\ell,\ell'=\lceil \beta \rceil}^\infty |c_{\ell'}| A_{\ell'}\left[\frac{{{\ell-\beta}}}{E_\ell +mc^2}-\frac{{\ell'-\beta}}{E_{\ell'} +mc^2} \right] A_{\ell} |c_{\ell}|\,\text{Sin}\left(\gamma_\ell-\gamma_{\ell'}\right) \text{Sinc}\left[\frac{(E_\ell-E_\ell')T}{2\hbar}\right] 
\,\,
\label{Eq:Prpsi}
\end{equation}

\end{widetext}

and  vanishes identically when all the coefficients $c_\ell$ are real i.e. $\gamma_\ell =0$ or $\pi$. The fact that with real coefficients the integral  of $J_r(0,t)$, over a symmetric time interval, vanishes is to be compared with what happens for the  corresponding  azimuthal component of the current which defines $P_\Psi$, defined in Eq.~\eqref{Eq:Ppsi}, whose absolute value is instead maximised in  the quantum backflow analysis by a state characterised by real coefficients.  
We notice that in \cite{PhysRevA.107.032204} the authors discuss the azimuthal quantum backflow of a non relativistic particle confined in a disk of radius $R$ punctured in the center by a magnetic flux line perpendicular to the disk, i.e. the particle moves in a Aharonov-Bohm like potential. 
We point out that the azimuthal backflow analysis in this truly two-dimensional system is carried out notwithstanding the presence of a radial current~\cite{PhysRevA.107.032204}. This radial current, while vanishing for the eigenstates, is  \emph{non vanishing} for a general superposition with complex coefficients similarly to what happens in the analysis of ref.~\cite{strange} and  in the present work as well.

\end{document}